\documentclass[twocolumn,final]{svjour3}   
\smartqed  

\AtBeginDocument{
\setlength{\oddsidemargin}{\dimexpr(\paperwidth-\textwidth)/2-1in}
\setlength{\evensidemargin}{\oddsidemargin}
\setlength{\topmargin}{\dimexpr(\paperheight-\textheight)/2-\headheight-\headsep-1in}}

\usepackage{graphicx}
\usepackage{amssymb,amsmath}
\usepackage[latin9]{inputenc}
\usepackage{multirow}

\begin{document}

\title{Towards a $^{229}$Th-based nuclear clock}

\author{Lars von der Wense \and
        Benedict Seiferle  \and
        Peter G. Thirolf
}

\institute{L. von der Wense \at
              Am Coulombwall 1, 85748 Garching, Germany \\
              Tel.: +49-(0)89-28914093\\
              \email{L.Wense@physik.uni-muenchen.de}           
           \and
           B. Seiferle \at
              Am Coulombwall 1, 85748 Garching, Germany
           \and
           P.G. Thirolf \at
           		Am Coulombwall 1, 85748 Garching, Germany
}

\date{}
\journalname{Measurement Techniques}
\maketitle
\vspace{-5cm}

\begin{abstract}
An overview over the current status of the development of a nuclear clock based on the state of lowest known nuclear excitation energy in $^{229}$Th is presented. The text is especially written for the interested reader without any particular knowledge in this field of research. It is thus ideal as an introductory reading to get a broad overview over the various different aspects of the field, in addition it can serve as a guideline for future research. An introductory part is provided, giving a historic context and explaining the fundamental concept of clocks. Finally, potential candidates for nuclear clocks other than $^{229}$Th are discussed.
\end{abstract}

\section{Introduction}
\label{intro}
\subsection{The history of time measurement}
During the evolution of time keeping devices, timekeepers were developed which technologically differ from each other in the maximum imaginable way. However, despite these differences, the abstract underlying principle remained unchanged. What is required for all clocks is a physical package that exhibits oscillatory or continuous changes and acts as a time giving device. Further, some mechanism is used which counts the number of oscillations or measures the changes and translates them into a time value.\\[0.2cm]
Already in historic cultures the measurement of time was an important tool for science and society and any improvement in the ability of time measurement was followed by technological advances. In ancient times mostly shadow clocks and sundials were used, both taking advantage of the highly stable earth rotation for time measurement, however revealing only limiting read-out precision \cite{Higgins}. Later, water clocks were developed and used in parallel to sundials. Here the constant flow of water is the time-giving process that allowed for higher read-out precision which is independent of the movement of celestial bodies, however also less stable and accurate than the earth rotation, thereby leading to larger inaccuracies than sundials. At the same time also fire clocks were used in China and candle clocks in European monasteries. First mechanical clocks were developed in the 13th century using a foliot in combination with a verge escape for time measurement and operated mostly in churches. These clocks were inaccurate by at least 15 minutes per day and required adjustment on a daily basis \cite{Higgins}.\\[0.2cm]
The first accurate mechanical clocks were pendulum clocks as developed by Christian Huygens in 1656, providing an error of less than 1 minute per day and later less than 10 seconds per day \cite{Bennet,Sorge}. In 1675 Huygens also developed the first portable pocket watches that were based on the spiral spring mechanism. For centuries, pendulum clocks remained the most accurate clocks and were constantly developed. Important improvements were implemented by John Harrison, who invented a pendulum clock achieving 1 second accuracy in 100 days and who also solved the longitude problem by developing the most accurate marine chronometer \cite{Gould}. The most accurate pendulum clock ever built is the Shortt-Synchronome free pendulum clock developed in 1921 \cite{Bosschieter}. It achieves an accuracy of about 1 second per year and was the first clock being more accurate than the earth rotation itself. Between the 1920s and 1940s, the Shortt clocks were operated in many countries as the most precise timekeeping devices.\\[0.2cm]
Pendulum clocks were outperformed in the 1940s by quartz clocks. The first quartz clocks, built already in 1927 by Joseph Horton and Warren Marrison, provided an accuracy of 10$^{-7}$, corresponding to about 1 second in 4 month and therefore significantly less than the Shortt pendulum clock \cite{Marrison}. The oscillation of a quartz clock is based on the piezo-electric effect and no macroscopic mechanical motion is required for timekeeping, which is a technological advantage over mechanical pendulum clocks and led to accuracies approaching 10$^{-9}$ in the 1940s \cite{Marrison}. This accuracy was much better than even the best mechanical pendulum clocks that were for this reason mostly substituted by quartz crystals.\\[0.2cm]
The age of the atomic clocks began in 1949, when Harold Lyons built the first ammonia maser at the National Bureau of Standards in the US \cite{Lyons}. Developing an atomic clock had already been suggested 4 years earlier by Isidor Rabi \cite{Forman}. In an atomic clock the frequency of light that corresponds to a certain energy difference between atomic shell states is used for time measurement. The first atomic clock was a proof-of-principle device and less accurate than the quartz clocks at that time. However, improvements led to the development of an accurate cesium standard by Louis Essen and Jack Parry in 1955 at the National Physical Laboratory in the UK, providing an accuracy of $10^{-9}$ \cite{Essen}. Due to technical improvements, the accuracy of atomic clocks enhanced significantly in the following years, approaching $10^{-13}$, corresponding to 1~s in 300,000 years in the mid 1960s \cite{Ramsey}. This led in 1967 to the decision to redefine the second to be 9,192,631,770 times the period of the $^{133}$Cs ground-state hyperfine transition. Previously the second was still defined based on the earth rotation. A further important improvement in time measurement was made in the end of the 1980s, when laser cooling allowed for longer interrogation cycles of the atomic beams, leading to the development of atomic fountain clocks with accuracies in the $10^{-16}$ range \cite{Wynands}.\\[0.2cm]
The next technological step was largely driven by the development of the frequency comb in 1998, that allows for direct measurement of frequencies in the optical range, five orders of magnitude larger than previously possible \cite{Udem}. This improvement was immediately applied for the development of single-ion optical clocks. The first clock of this type was presented in 2001 and was a single-ion $^{199}$Hg$^{+}$ clock, providing a fractional frequency instability of $7\cdot10^{-15}$ measured in 1 second of averaging \cite{Diddams}. In the meantime a multitude of single-ion optical clocks were developed and are operated in metrology institutes around the world \cite{Ludlow}. A 2008 comparison of two single-ion optical clocks revealed a fractional frequency uncertainty of $5.2\cdot10^{-17}$ \cite{Rosenband} and in 2016 the operation of a single ion Yb$^+$ optical clock at $3\cdot10^{-18}$ frequency uncertainty was reported \cite{Huntemann}. In parallel, also optical lattice clocks were developed. These clocks are based on neutral atoms instead of ions, providing the advantage that no Coulomb interactions between individual particles are present, which results in high achievable densities when atoms are stored in an optical lattice, thereby significantly increasing statistics. By today optical lattice clocks are the most accurate clocks in the world, approaching a total frequency uncertainty of $10^{-18}$ \cite{Bloom,Nicholson}\footnote{Addendum: In 2018 an Yb optical lattice clock with a total frequency uncertainty of $1.4\cdot10^{-18}$ was reported \cite{Mcgrew}.}.

\subsection{Quality of clocks}
Independent of the exact process of time measurement, there must be way how to compare the quality of different clocks. The quality of a clock will depend on various factors, however, there are two quantities which play the most important role when discussing the quality of clocks: Accuracy and stability. Quite generally, the accuracy describes how much a measured value differs from the correct value. In the context of time measurement, two different accuracies have to be distinguished: The accuracy of the actual time measurement and the frequency accuracy of the clock. The time accuracy (usually just referred to as "accuracy") expresses how much a clock deviates from the actual time and is given in relative units, e.g. if a clock deviates by 1 second from the correct time, 24 hours after it was synchronized with the correct time, the clock has an accuracy of 1 second per day, or $1.1\cdot10^{-5}$. The definition of accuracy requires the comparison with a time value that is said to be "correct". In contrast, the frequency accuracy (usually referred to as the "systematic frequency shift") is a measure for the frequency deviation of a real oscillator to the ideal value. If the ideal frequency of an oscillator is given as $\omega_0$, the real oscillator will most likely exhibit a systematic shift leading to a real frequency $\omega_\text{real}=\omega_0(1+\epsilon)$. Here $\epsilon$ denotes the systematic frequency shift.\\[0.2cm]
If many clocks of the same type were compared, each clock can be expected to possess a slightly different systematic frequency shift, leading to a distribution of frequency shifts. The size of this distribution will be the larger, the more sensitive the clock reacts to external perturbations. For this reason, the uncertainty in $\epsilon$, usually referred to as the "systematic frequency uncertainty" is an important parameter for the quality of a clock. In case that the ideal oscillator value $\omega_0$ would be set as a standard used to define the correct time, the systematic frequency uncertainty would be a measure for the time accuracy that could on average be achieved with the type of oscillator. This is the case, because one can correct for the systematic frequency shift $\epsilon$ up to its uncertainty. The remaining value will then lead to an unavoidable deviation of the frequency used for time measurement to the ideal frequency and thus to a continuously growing error in time measurement, which equals the time accuracy. For this reason, the total systematic frequency uncertainty is often referred to as the clock's accuracy.\\[0.2cm]
The stability of a clock is a measure for statistical fluctuations that occur during the frequency measurement. If the ideal frequency is denoted with $\omega_0$, the measured frequency will in general be time dependent in accordance with $\omega_\text{meas}(t)=\omega_0(1+\epsilon+y(t))$. Here $y(t)$ is called the frequency noise and the variance of $y(t)$ is a measure for the stability of a clock. The variance is given as the Allan variance $\sigma^2_y(\tau)$, which is defined as the comparison of two successive frequency deviations for a given sampling period $\tau$. More exactly, we have $\sigma^2_y(\tau)=1/2\langle\left(\bar{y}_{n+1}-\bar{y}_n\right)^2 \rangle$, with $\bar{y}_n$ as the $n$th fractional frequency average over the sampling period $\tau$. For current atomic clocks, the best Allan deviations $\sigma_y(\tau)$ that can be achieved are limited by the quantum projection noise (QPN) and are given by \cite{Ludlow}
\begin{equation}
\label{Allan1}
\sigma_y(\tau)\approx \frac{1}{Q}\sqrt{\frac{T}{N\tau}}.
\end{equation}
Here $Q=\omega/\Delta\omega$ is the quality factor of the resonance, $T$ the time of the interrogation cycle (which cannot exceed $1/\Delta\omega$), $N$ the number of irradiated atoms and $\tau$ the sampling period. For the most stable atomic lattice clock, an Allan deviation of $2.2\cdot10^{-16}/\sqrt{\tau}$ for $1$~s interrogation cycle and thus close to the QPN limit was reported \cite{Nicholson}. A total frequency uncertainty of $2.1\cdot10^{-18}$ was achieved. The clock parameters for this strontium lattice clock are $Q=2\cdot10^{14}$ and $N\approx 2000$. The best single-ion clock currently in operation ($^{171}Yb^{1+}$ E3) provides a total frequency uncertainty of $3\cdot10^{-18}$ and an Allan deviation of $5\cdot10^{-15}/\sqrt{\tau}$ ($T=1$~s, $Q\approx10^{23}$, $N=1$) \cite{Huntemann}. The calculated QPN limit is $10^{-23}$, significantly below the obtained value. The difference originates from an extremely small natural linewidth of the transition in the nHz region, which is drastically smaller than the bandwidth of any laser available for excitation. For this reason the stability is not limited by the transition linewidth, but instead by the bandwidth of the laser light and only limited advantage is gained from the extremely small natural linewidth of the transition.

\section{The idea of a nuclear clock}

\subsection{Advantages of a nuclear clock}
Although the accuracy that is achieved by the best atomic clocks today is already stunning, approaching $10^{-18}$ corresponding to an error of 1 second after $3\cdot10^{10}$~years, significantly longer than the age of the universe, it is reasonable to ask if it is possible to push the limits further. The performance of optical lattice clocks is by today mostly limited due to the influence of external perturbations like electric and magnetic fields. A natural way of improving clock performance would be to use nuclear transitions instead of atomic shell transitions for time measurement. Conceptually, nuclear transitions provide three advantages (see also Ref.~\cite{Peik2}): 1. The atomic nucleus is about 5 orders of magnitude smaller than the atomic shell, which leads to significantly reduced magnetic dipole and electric quadrupole moments and therefore to a higher stability against external influences, resulting in an expected improved accuracy and stability of the clock. 2. The transition frequencies are larger, thereby in principle allowing for small Allan deviations. 3. The nucleus is largely unaffected by the atomic shell, for this reason it is intriguing to develop a solid-state nuclear clock based on Mössbauer spectroscopy. Such a solid-state clock could contain a high density of nuclei of about $10^{14}$~mm$^{-3}$, thus leading to much improved statistical uncertainties when compared to atomic lattice clocks providing about $10^4$ atoms.\\[0.2cm]
Generally, two different classes of nuclear clocks can be considered: (1) Solid-state nuclear clocks that make use of laser-based Mössbauer spectroscopy and would have the advantage of a high density of nuclei, leading to high statistics, but may eventually suffer from line broadening effects and (2) single-ion nuclear clocks, which provide drastically lower statistics, but for which the environmental conditions could be by far better controlled. The operational principles very much resemble those of atomic optical lattice clocks and atomic single-ion clocks, except that the atomic transition is replaced by a nuclear transition.

\subsection{$\gamma$ decay versus internal conversion}
There is one difference which slightly complicates the discussion of nuclear clocks compared to atomic clocks: While for atomic transitions only one decay channel of the excited states enters the discussion, namely radiative decay via emission of a photon, nuclear transitions may exhibit more relevant decay channels. The most important ones in this context are radiative decay ($\gamma$ decay) and internal conversion (IC). During the internal conversion process the nucleus couples to the atomic shell, transferring the excited state's energy to an electron, which is subsequently emitted into the continuum. The fraction of IC compared to $\gamma$ decay is called the internal conversion coefficient and usually denoted as $\alpha_\text{ic}$. Typically, IC is even the dominant decay channel of excited nuclear states with low energy, long lifetime and large atomic number. The possibility of IC leads to a shortening of the excited state's lifetime and thereby to a line broadening, which is unwanted when considering clock performance. Also care has to be taken when calculating nuclear excitation probabilities, as only the linewidth corresponding to the radiative decay rate enters the calculation. Thus nuclear states that appear to have a rather short lifetime might still have a negligible excitation probability if the internal conversion coefficient is large.\\[0.2cm]
One possibility to reduce the IC decay branch of a nuclear transition is to increase the charge state of the considered ion. For nuclear excited states of low energy, there exists a certain charge state for which the ionization potential of the ion starts to exceed the energy of the nuclear state. If this is the case, the excited state's decay does not provide sufficient energy to lead to further ionization and the IC decay is energetically forbidden. Note, that this will lead to an increased lifetime of the nuclear state and, correspondingly, to a smaller linewidth of the transition. It will, however, not result in an enhanced excitation probability of the nuclear state, as this value is connected to the radiative decay rate, which remains unaffected from a reduced IC decay rate. Even when considering low-energy nuclear states, the suppression of the IC decay channel would typically require the use of highly-charged ions. One example is the low-energy state $^{205\text{m}}$Pb (2.3~keV, 24.2~$\mu$s, E2), where the lifetime of 24.2~$\mu$s is rather low considering a potential application for a nuclear clock. The IC coefficient is however large, with a value of $\alpha\approx 4.4\cdot10^8$ and from Pb$^{46+}$ onwards the IC decay channel can be expected to be energetically suppressed, leading to a drastic prolongation of the lifetime to up to $10^4$~s, thus in principle providing a good candidate for a highly-charged-ion nuclear clock. For solid-state nuclear clocks, the ionization potential has to be replaced by the material's band gap, however, there is practically no overlap of the energy range of band gaps with the nuclear energy scale. For this reason there is no hope for the suppression of the IC decay channel in the solid-state nuclear clock concept ($^{229\text{m}}$Th provides an exception in this context).

\subsection{What makes $^{229\text{m}}$Th unique}
By today, there is a major obstacle that prevents the development of a broad variety of nuclear clocks. This is that even energies of several keV, corresponding to wavelengths of 1~nm and below, are considered as low-energy transitions on nuclear energy scales and by today the frequency-comb technology has not been sufficiently extended into the x-ray region in order to drive most of the nuclear transitions. As the frequency comb technology is a young and fast improving field of technology, one might expect that the availability of frequency combs in the energy range of a few keV with sufficient intensities to drive nuclear transitions and thus to allow for high-precision Mössbauer spectroscopy is just a matter of time. In the meantime, however, the research on nuclear clocks has to focus on a single nuclear state providing an extraordinary low energy of only about 7.8~eV above the nuclear ground state. This is the first excited state of $^{229}$Th, denoted as $^{229\text{m}}$Th, where the superscript m stands for metastable and expresses that a comparatively long lifetime for this state is expected. To give an impression on the scarcity of nuclear states of such low energy: There is only one further nuclear state known with an energy of less than 1~keV above its nuclear ground state, which is $^{235\text{m}}$U, providing an energy of about 76~eV. This isomeric state, however, does not allow for the development of a nuclear clock, due to its extremely long radiative lifetime of $3.8\cdot10^{22}$~s, which is about 100,000 times the age of the universe, not allowing for any significant laser coupling. For this reason, $^{229\text{m}}$Th is by today the only realistic candidate for the development of a nuclear frequency standard. In the following only $^{229\text{m}}$Th will be considered. A discussion of future nuclear clock candidates is provided in the outlook.

\section{$^{229\text{m}}$Th for time measurement}

\subsection{Early research on $^{229\text{m}}$Th}
In the 1970s, detailed measurements of the $\gamma$-ray spectrum accompanying the $^{233}$U $\alpha$ decay were performed. $^{233}$U is a radioactive isotope with a half-life of 159,000~years, decaying to $^{229}$Th. The measured spectra revealed some features which could only be explained if the existence of an excited nuclear state in $^{229}$Th close to the ground state was assumed. The decay of this excited state to the ground state could, however, not be observed, which was attributed to a low excitation energy. The excitation energy was then inferred to be less than 100~eV, purely based on the non-observation of the ground state decay in 1976 \cite{Kroger}. While this discovery generated only little excitement at that time, measurements that were aiming for improved energy constraints continued. In 1990 the energy was constrained to $-1\pm4$~eV, using differences of $\gamma$ rays of higher energy, which led to the conclusion that the energy must be almost certainly below 10~eV \cite{Reich}. This measurement was further improved, resulting in an energy value of $3.5\pm1$~eV published in 1994, which was for a long time the most accepted energy value \cite{Helmer}.\\[0.2cm]
In 2007, however, a new result of an indirect energy study of $^{229\text{m}}$Th was published by Beck et al., leading to a corrected energy value of $(7.6\pm0.5)$~eV \cite{Beck1}. In their work they used a newly available microcalorimetric x-ray spectrometer with a resolution of $\sim30$~eV for $\gamma$-ray spectroscopy of the $^{233}$U decay. The isomeric energy value was further improved in 2009 to $(7.8\pm0.5)$~eV by the same group \cite{Beck2}. This remains until today to be the most accepted value for the isomeric energy.\\[0.2cm]
The correction of the energy value from 3.5~eV to 7.8~eV allows to explain why early experiments had failed to observe any direct isomeric decay signal. Firstly, an energy of 7.8~eV corresponds to a wavelength of about 159~nm emitted in the decay to the ground state, which is deep in the vacuum-ultraviolet region and poses special requirements for the applied optical system. But even more importantly, the new energy value is above the first ionization potential of thorium of 6.3~eV, making internal conversion to become the dominant decay channel of the isomeric state in neutral thorium atoms. A calculation performed by F.F. Karpeshin and M.B. Trzhaskovskaya in 2007 revealed that an internal conversion factor of $\alpha\approx10^9$ has to be expected for the isomeric state in neutral thorium, leading to a lifetime shortening by the same factor to about 10~$\mu$s \cite{Karpeshin}. All except one experiments at that time had searched for a radiative decay of the isomeric state in neutral thorium atoms, which can be considered has hopeless, given the corrected energy value (see Ref.~\cite{Wense1} for a detailed review). The only experiment that had searched for an IC decay channel was performed in 1995 by O.V. Vorykhalov and V.V. Koltsov and was not sensitive to short lifetimes \cite{Vorykhalov}. For this reason it was not surprising that searches for a direct observation of the nuclear transition had been unsuccessful and a new series of experiments was started, aiming for a direct detection of $^{229\text{m}}$Th.\\[0.2cm]

\subsection{The $^{229}$Th nuclear clock proposal}
The observation of a nuclear excited state of only a few eV above the ground state immediately attracted some attention from the scientific community. Already in 1991 a theory paper appeared, predicting an increasing interest of physicists from other disciplines like {\it "optics, solid-state physics, lasers, plasma and others"}, different decay channels and lifetimes were discussed in dependence on the transition energy \cite{Strizhov}. In a 1996 paper of E.V. Tkalya et al. a list of potential applications is provided, which also contains the potential for the {\it "development of a high stability nuclear source of light for metrology"} \cite{Tkalya1}. While this proposal does not include the development of a nuclear clock based on $^{229}$Th, it contains a hint pointing at an expected high stability. A few years later also the decay of $^{229\text{m}}$Th in solid state materials was discussed \cite{Tkalya2}.\\[0.2cm]
On this basis, a  $^{229\text{m}}$Th nuclear clock was proposed in 2003 by E. Peik and C. Tamm, shortly after the development of the frequency-comb technology, which for the first time allowed precision laser spectroscopy in the optical region \cite{Peik}. This paper contains both nuclear-clock concepts: A clock based on Th ions as well as a solid state nuclear clock.\\[0.2cm]
$^{229\text{m}}$Th, besides being an excited nuclear state of extraordinary low energy, possesses a lifetime corresponding to its radiative decay channel of expectedly about $10^4$~s \cite{Minkov,Tkalya3}. This leads to a large quality factor of the corresponding resonator of $\omega/\Delta\omega\approx10^{20}$ and a small absolute bandwidth of $10^{-4}$~Hz in case that IC is suppressed. The ground and isomeric states have spin and parity values of $5/2^+$ and $3/2^+$, respectively, leading to a multipolarity of M1 of the isomer-to-ground-state transition, which results in a sufficiently large cross section to allow for direct laser excitation. Further, the ground state of $^{229}$Th has a half-life of 7932 years, which makes it relatively easy to handle moderate quantities of the material. All properties are requirements for the development of a $^{229}$Th-based nuclear clock.\\[0.2cm]
In their 2003 nuclear clock concept Peik and Tamm proposed to perform nuclear laser spectroscopy with $^{229}$Th$^{3+}$ \cite{Peik}. The $3+$ charge state was proposed because it possesses a favorable electronic configuration with a radon-like core and one valence electron, exhibiting a closed 3-level $\Lambda$ system and a closed 2-level system, which can be employed for laser cooling. Due to the small magnetic dipole and electric quadrupole moments of the nucleus, the direct coupling of external perturbing fields to the nucleus are negligible. However, shell-nucleus coupling via the hyperfine interaction still has to be considered as a potential source of perturbations. One of the central ideas of the 2003 nuclear-clock concept is that one could choose an excited state of the electronic shell in a way that the combined quantum numbers of shell plus nucleus are "good" in a sense that the entire system provides lowest sensitivity to external perturbations \cite{Peik2}. It was shown that the metastable $7s^2S_{1/2}$ shell state in Th$^{3+}$ with 1~s lifetime would be an appropriate choice. In this way, perturbing effects due to the linear Zeeman effect, the tensor part of the quadratic Stark effect and atomic quadrupole interactions could be avoided. Further, as no shifts can play a role which are entirely dependent on the electronic quantum numbers, no shifts from static electric fields, electromagnetic radiation or collisions have to be considered, leading to the proposal of a highly stable nuclear clock. The population of the nuclear excited state could be monitored by making use of the change of the nuclear spin state, resulting in a different hyperfine structure of the cooling transitions (double-resonance method). It was also suggested that even a solid-state nuclear clock could be realized, if $^{229}$Th was embedded into a material providing a band gap larger than the nuclear isomer's energy, for which $(3.5\pm1)$~eV was the assumed value at that time \cite{Peik2,Peik}.\\[0.2cm]
This pioneering nuclear clock concept, besides promising to lead to an extraordinary stable clock, has two disadvantages \cite{Campbell}: (1) The quadratic Zeeman effect cannot be suppressed and is estimated to be 1~kHz at 0.1~mT, which is comparable to usual atomic clocks. (2) The concept requires to excite the $^{229}$Th$^{3+}$ ions into the $7s^2S_{1/2}$ atomic shell state, which possesses 1~s lifetime (instead of the expectedly $\sim10^4$~s lifetime of the nuclear isomer) and will thus lead to a reduced quality factor of the resonance.\\[0.2cm]
A more recent proposal by C. Campbell et al. aims at a solution of these problems \cite{Campbell}. Here the idea is to use a pair of stretched nuclear hyperfine states for the clock transition, while $^{229}$Th$^{3+}$ remains in its $5f^2F_{5/2}$ electronic ground state. A careful analysis of the expected systematic uncertainties of a corresponding $^{229}$Th$^{3+}$ single-ion nuclear clock was performed, resulting in an expected total uncertainty approaching $10^{-19}$, thereby potentially outperforming all existing atomic-clock technology \cite{Campbell}. The development of a nuclear clock based on $^{229}$Th ions is currently envisaged by three groups in the world, with more eventually to follow \cite{Campbell3,Zimmermann,Borisyuk}.\\[0.2cm]
The concept of a solid-state nuclear clock based on Mössbauer spectroscopy of $^{229\text{m}}$Th, as first proposed in Ref.~\cite{Peik}, was further developed by two groups \cite{Rellergert,Kazakov1}. A long-term fractional frequency accuracy of $2\cdot10^{-16}$ has been predicted for this device \cite{Rellergert}, with a fractional instability approaching $10^{-19}$ \cite{Kazakov1}.

\section{Constraining the transition energy}
The development of a $^{229}$Th-based nuclear clock requires direct laser excitation of the isomeric nuclear state. However, due to the narrow radiative bandwidth of the nuclear transition of expectedly only $10^{-4}$~Hz, the laser-nucleus interaction is weak and improved constraints on the transition energy are required as a prerequisite for the nuclear laser excitation of $^{229}$Th ions in a Paul trap. Up to today it was not possible to constrain the isomeric energy value sufficiently in order to allow for the development of a nuclear clock. For this reason major experimental investigations are continued, aiming for a reduction of the uncertainty of the $^{229\text{m}}$Th energy value. Except for a few experiments, most of the experimental concepts are making use of a direct decay detection of the isomer-to-ground-state transition for constraining the transition energy. As the isomeric decay can occur via two decay channels, experiments that are aiming for a direct detection of $^{229\text{m}}$Th can be divided into two groups: (A) Experiments aiming for the detection of a radiative decay channel and (B) experiments looking for an internal conversion (IC) decay channel of the isomeric state. Both of these groups can be further subdivided by the way chosen for the isomer's population, which could either be done by (1) a natural decay branch from a mother nuclide or (2) an excitation from the $^{229}$Th ground state. Besides that, there are a few experimental concepts planning to determine the energy via indirect methods (C). All types of experiments will be discussed individually in the following.

\subsection{Experiments searching for a radiative decay}
\subsubsection{Isomer population from a natural decay branch}
There are two natural decay branches from mother nuclides into the $^{229}$Th isomeric state: One via the $\alpha$ decay of $^{233}$U, where the isomer is populated by 2\% and the other via the $\beta$ decay of $^{229}$Ac, which populates the isomeric state to expectedly 13.4\%. Most of the experiments that have so far been conducted are making use of the 2\% population branch from the $^{233}$U $\alpha$ decay. The dominant reason is that $^{229}$Ac has a half-life of about 63~min only (as opposed to $^{233}$U with a half-life of $1.6\cdot10^5$ years), and has to be continuously produced with the help of accelerators. Besides that, the $^{233}$U $\alpha$ decay transfers larger momentum to the $^{229}$Th nucleus, thereby allowing to separate the $^{229}$Th recoil nuclei from the source material, which provides a major advantage for background reduction.\\[0.2cm] 
Several types of experiments were performed, in which the 2\% decay branch in the $\alpha$ decay of $^{233}$U was employed in order to populate the $^{229}$Th isomeric state. In one class of these experiments, $^{229}$Th $\alpha$-recoil ions, emerging from a large-area $^{233}$U source, are implanted into a catcher foil consisting of a large band-gap material. As the IC is expected to be suppressed due to the large band gap, the isomeric decay is probed in the radiative decay channel. In case of success, the isomeric energy could be probed via optical filters or VUV spectrometers. Experiments along this line were performed at the PTB in Germany \cite{Zimmermann}, at the Lawrence Livermore National Laboratory (LLNL) in the US \cite{Swanberg} and at the Los Alamos National Laboratory (LANL), US \cite{Zhao}. Only the last group reported a successful direct observation of $^{229\text{m}}$Th, however, the result has been stated doubtful \cite{Peik3}. Problems of this approach are a potentially large non-radiative decay channel and strong background signals originating from radioluminescence.\\[0.2cm] 
An improved proposal based on the same concept was presented by our group at the Ludwig-Maximilians-University Munich, Germany in 2013 \cite{Wense2}. Here it was the idea to thermalize the $^{229}$Th $\alpha$-recoil ions originating from the $^{233}$U decay with the help of buffer gas and to form an isotopically pure $^{229}$Th ion beam, before implanting the ions into a large band-gap material. This concept has the advantage of a drastic reduction of background radioluminescence and an increased signal-to-background ratio of expectedly up to $10^4$, due to a nearly point-like implantation region, which allows for the application of a highly efficient optical focusing system. In case of a successful observation of light emitted from the isomeric decay, the energy could have been probed by optical filters or via spectroscopy. However, no radiative decay signal was observed, which can clearly be attributed to an unsuccessful suppression of the IC decay channel \cite{Wense1}.\\[0.2cm]
A different approach was proposed in 2013 at LANL and further developed at the TU Vienna \cite{Hehlen,Stellmer3}. Here it is the central idea to grow $^{233}$U into a suitable host crystal and $^{229}$Th is implanted from the $^{233}$U $\alpha$ decay directly into the bulk material of the crystal. It is searched for a potential radiative decay channel of the isomeric state. The energy could again be probed with the help of filters or spectrometers. Experiments along this line are still in progress, however, so far no signal originating from $^{229\text{m}}$Th could be observed. Potential reasons are a large non-radiative decay channel and background originating from radioluminescence.\\[0.2cm]
Very recently, a proposal was put forward to also investigate the 13.4\% decay channel from the $\beta$ decay of $^{229}$Ac \cite{Duppen}. It is the idea to produce $^{229}$Ac in form of an isotopically pure ion beam at ISOLDE (CERN) and implant the ions into a large band-gap material. The $^{229}$Ac lifetime of 63~min allows to anneal the material to make sure that actinium is properly embedded into the crystal lattice. Due to the low momentum transferred in the $\beta$ decay, $^{229}$Th will occupy the same place in the crystal and the IC decay channel of the isomer is expected to be suppressed, allowing to detect the radiative isomeric decay channel with the help of an efficient focusing system, as soon as the $^{229}$Ac activity has sufficiently faded away. In case of successful observation of the radiative decay, the isomeric energy could be determined by making use of filters or VUV spectroscopy. It can be expected that a non-radiative decay channel and background radioluminescence caused by the $^{229}$Ac $\beta$ decay will pose major challenges to this concept.

\subsubsection{Isomer population from the $^{229}$Th ground state}
While the direct nuclear laser excitation of $^{229\text{m}}$Th ions in Paul traps is not yet feasible, a different concept for laser excitation of the isomeric state was proposed, making use of the so-called inverse electronic bridge (IEB) effect \cite{Porsev}. In the IEB process, an atomic shell state is excited by laser light and couples to the nucleus. The nucleus is then excited with some probability during the decay of the atomic shell state. Light, as subsequently emitted during the nuclear isomeric decay, should be observable. Experiments along this line have been performed at the PTB in Germany \cite{Herrera}, at Georgia Tech in the US \cite{Campbell2,Campbell3} and are in preparation in the framework of the Moscow collaboration in Russia \cite{Borisyuk}. In the Georgia Tech experiment it is possible to also detect the isomeric state's population via the change of the hyperfine structure, which can be probed by the cooling lasers \cite{Campbell3}. A more precise determination of the energy could be obtained either by using filters or by varying the frequency of the laser used for electron excitation. The experiments have so far been unsuccessful, major obstacles are uncertain IEB nuclear excitation rates, a broad energy range required to be scanned and low expected signal rates.\\[0.2cm]
A different approach for the determination of the $^{229\text{m}}$Th energy was proposed in 2010 \cite{Rellergert}. Here the idea is to dope large band-gap materials with a high concentration of $^{229}$Th. The obtained crystals are then irradiated with broad-band light sources to excite the nuclear transition. It is searched for light as expected to be emitted during the isomeric decay. The achievable $^{229}$Th densities of above $10^{13}$~mm$^{-3}$ are sufficient to lead to significant absolute numbers of excited nuclei even for moderate spectral energy densities of the light source. In case of a successful detection, the energy could be probed with the help of filters or VUV spectrometers. Experiments along this line were performed by the University of California in the US \cite{Jeet} and are in preparation at the Technical University of Vienna, Austria \cite{Stellmer}\footnote{Addendum: Recently a first measurement from this group was reported \cite{Stellmer4}.}. Unfortunately, no light emitted from the isomeric decay could so far be observed. The reason is most likely a significant non-radiative decay branch, which leads to a loss of signal intensity. Major challenges are photoluminescence and radioluminescence effects of the crystals.\\[0.2cm]
Results of a similar experiment, in which $^{229}$Th was not deposited in a crystal but instead on the surface of a large band-gap material and irradiated with synchrotron radiation, were reported \cite{Yamaguchi}. No light signal originating from the isomeric decay could be observed, which is most likely due to an unsuccessful suppression of the IC decay channel.\\[0.2cm]
It was also proposed to excite the 29~keV state of $^{229}$Th in doped crystals by synchrotron radiation \cite{Yoshimi}. This state decays into $^{229\text{m}}$Th, for which a subsequent photonic decay should be observable. A similar experiment was already reported in 2005 \cite{Gangrsky}. Photoluminescence and radioluminescence as well as loss of isomeric signal due to non-radiative decay channels will pose major challenges.

\subsection{Experiments making use of IC decay}
\subsubsection{Isomer population from a natural decay branch}
The experiments described above are aiming for an observation of the isomeric decay in the radiative decay channel. However, since 2007 it is known that the isomeric state will strongly (by a factor of about $10^9$) favor the IC decay if energetically allowed \cite{Karpeshin}. Any detection of a photon emitted during the isomer's deexcitation will therefore require the suppression of the IC decay by a factor of about a billion, which can be expected to pose a considerable experimental challenge. Instead of trying to suppress the IC decay, it is intriguing to use this channel for probing the isomeric decay. Surprisingly few experiments were following this path of research, although already proposed in 1991 \cite{Strizhov}.\\[0.2cm]
One early experiment aiming for the detection of an IC decay channel was performed in Russia and published in 1995, but was not sensitive for short lifetimes \cite{Vorykhalov}. More recently, a class of experiments was performed at LLNL in the US, thereby probing the huge lifetime range of 13 orders of magnitude between $4\cdot10^{-8}$~s and $2\cdot10^5$~s \cite{Swanberg}. In one of these experiments, $^{229}$Th $\alpha$-recoil ions were implanted into a catcher foil and electrons, emitted after the implantation, were observed by an MCP detector. The detection was triggered in accordance with the $^{233}$U $\alpha$ decays. Although providing the potential for the direct detection of $^{229\text{m}}$Th, no IC decay signal was observed. The reason is subject to speculation and might be a too deep implantation depth of the $^{229}$Th recoil ions, thereby hindering the IC electrons from efficiently leaving the catcher foil\footnote{Addendum: Results of a similar experiment were reported in 2018 \cite{Stellmer5}. Also here no electrons originating from the isomeric decay could be securely identified.}.\\[0.2cm]
Only recently our group changed the direction of research from the observation of a radiative decay channel, as originally aimed for but proven unsuccessful \cite{Wense1,Wense2}, to the investigation of an IC decay. In these experiments we were making use of our isotopically pure, low-energy $^{229}$Th ion beam as produced from the $\alpha$ decay of $^{233}$U by thermalization in a buffer gas. Instead of implanting the ions into a large band-gap crystal, we accumulated them with low kinetic energy directly on the surface of an MCP detector. During this process the ions neutralize, thereby triggering the isomeric IC decay. Subsequently, the low-energy electrons emitted in this process are detected by the MCP detector. This experiment has finally led to the direct detection of the $^{229\text{m}}$Th decay \cite{Wense3}. Based on this direct decay signal, the $^{229\text{m}}$Th half-life in neutral thorium could be determined to be about 7~$\mu$s \cite{Seiferle1}, well in agreement with the theoretically predicted value \cite{Karpeshin}. It is now envisaged to determine the isomer's energy by conversion-electron spectroscopy \cite{Wense3}. By making use of a magnetic-bottle type retarding-field spectrometer, it should be possible to significantly pin down the isomeric energy value \cite{Seiferle2}. An alternative approach could be to use a microcalorimeter for energy determination \cite{Ponce}.

\subsubsection{Isomer population from the $^{229}$Th ground state}
To the best of our knowledge there is no evidence for experiments reported in literature that have aimed to excite the nuclear isomer from the $^{229}$Th ground state and to detect the isomer's deexcitation via the IC decay channel. Such experiments would form a whole new group, as the excitation could be performed in various ways. There are several ideas on how to excite the isomeric state from the $^{229}$Th ground state. The first idea is to use light for the isomer's direct excitation. Due to the low isomeric cross section, such experiments require a large number of nuclei as long as the energy has not been further constrained. Other ways that have been proposed for excitation of the isomer are: (1) Excitation in a laser-produced plasma \cite{Strizhov} (2) using the electronic shell as a bridge \cite{Tkalya1} (3) via excitation of the 29~keV nuclear state by synchrotron radiation, which will subsequently decay to the isomer \cite{Yoshimi,Gangrsky}, (4) by electron beam irradiation \cite{Borisyuk} and (5) via surface plasmons \cite{Varlamov}.\\[0.2cm]
An experiment of this type would even allow for direct laser excitation of $^{229\text{m}}$Th in a solid sample without the requirement of an improved knowledge of the isomer's excitation energy and has the potential to pin down the isomeric energy to a fraction of an eV. The experimental concept is to perform laser-based conversion electron Mössbauer spectroscopy (CEMS) by irradiating a thin $^{229}$Th layer with tunable and pulsed VUV laser light \cite{Wense4}. As the spectral energy density of individual laser pulses is relatively large, a significant number of nuclei could be excited into the isomeric state during each individual laser pulse. Under these conditions the isomeric decay would securely occur by IC with a half-life of 7~$\mu$s. The detection of the IC electrons emitted from the sample could be triggered in coincidence with the laser pulses in order to drastically improve the signal-to-background ratio. A calculation reveals that, using a laser of 10~GHz bandwidth for scanning, it could be possible to scan the large energy interval of 1~eV within a few days, obtaining a signal-to-background ratio of $10^4$ in case of resonance with the nuclear transition. Besides representing a first direct nuclear laser excitation scheme for $^{229\text{m}}$Th, such an experiment would allow to ultimately pin down the energy with sufficient precision to also allow for laser excitation of $^{229}$Th in Paul traps and thus for the development of a single-ion nuclear clock. Further, the concept could also allow for the development of a solid-state nuclear clock based on CEMS \cite{Wense1}. Experiments along this line are currently in preparation. 

\subsection{Indirect experiments for $^{229\text{m}}$Th energy determination}
There are few approaches that aim for an improved energy value without the requirement of a direct detection of the isomeric decay. One idea is to repeat the measurement which was performed by Beck et al. in 2007 and has led to the currently accepted energy value of ($7.8\pm0.5$)~eV \cite{Beck1,Beck2}. Within the past 10 years the technology of microcalorimeters has drastically evolved and spectrometers with an energy resolution of about 3~eV are now available \cite{Kazakov}. No conceptual restrictions for this experiment exist and the only challenge is to develop a calorimetric detector optimized for the nuclear excited state of 29~keV energy and providing sufficient resolution to resolve the closely-spaced doublet. A corresponding detector system is currently under development at the Kirchhoff-Institute for Physics in Heidelberg, Germany \cite{Schneider}.\\[0.2cm]
A different idea that does not depend on the direct detection of the isomeric decay is to excite the isomeric state via the IEB mechanism in a Coulomb crystal of $^{229}$Th$^{3+}$ ions. The isomer's excitation can then be probed via the hyperfine shift which is induced by a different nuclear spin state and the isomeric energy can be inferred to be close to the energy of the laser light used to trigger the IEB process. Experiments along this line were conducted at Georgia Tech, being the only place where Coulomb crystals of $^{229}$Th$^{3+}$ are available \cite{Campbell3}. No successful excitation of $^{229\text{m}}$Th has yet been reported.\\[0.2cm]
A further approach is investigated at storage rings. Here it is the idea to use the process of nuclear excitation by electron capture (NEEC), which is the inverse of the IC process, to infer the isomeric properties. $^{229}$Th ions are stored in a high-energy storage ring. When these ions catch electrons that fulfill the resonance condition (that the electron's kinetic energy plus the binding energy after recombination equals the energy of the isomeric state), there is an enhanced probability to excite $^{229}$Th into its isomeric state \cite{Palffy}. By tuning the energy of an electron beam and monitoring the number of recombinations by detecting the ions' charge states, it is possible to find the resonance and thus to determine the isomeric energy. Experiments are in preparation at GSI in Germany \cite{Brandau} and at the IMP in China \cite{Ma}.\\[0.2cm]
An experimentally more complex proposal to determine the isomeric energy via optomechanically induced transparency was recently proposed \cite{Liao}. A corresponding experiment can be expected to be technologically challenging. It was also proposed to use the hyperfine structure of $^{229}$Th$^{3+}$ in its nuclear ground state to probe the isomer's energy \cite{Beloy}. By today, no experiments are known to us that follow these proposals.

\section{Promising perspectives for future research}

\subsection{Steps towards a nuclear clock}
There are a few steps left on the way towards the development of a nuclear clock, these are: 1. The precise determination of the $^{229\text{m}}$Th energy value, 2. The measurement of the $^{229\text{m}}$Th hyperfine structure as a probe for the nuclear excitation\footnote{Addendum: This measurement has recently been performed \cite{Thielking}.}, 3. Direct laser excitation of $^{229\text{m}}$Th in a Paul trap.\\[0.2cm]
It is foreseeable that the first and most important step of a precise determination of the isomeric energy will come to a conclusion within the next few years. Experiments that will most likely succeed in this task are (1) an IC-electron spectroscopy experiment, which is based on the recent direct detection of the isomeric decay and currently conducted in our group \cite{Wense3,Seiferle2} and (2) an indirect approach, applying a state-of-the-art metallic magnetic microcalorimeter at the Kirchhoff Institute in Heidelberg \cite{Kazakov,Schneider}. For the success of the direct approach using IC-electron spectroscopy it is important to note that, opposed to a wide-spread assumption, this concept is independent of the work function of the material used for ion neutralization and can be applied for a precise energy determination \cite{Seiferle2}. Both concepts can therefore allow to constrain the energy to about 0.1~eV.\\[0.2cm]
In a second step, a direct laser excitation scheme based on conversion electron Mössbauer spectroscopy (CEMS) can be employed to ultimately pin down the isomeric energy value to below $10^{-6}$~eV, which corresponds to typical hyperfine energies in Mössbauer spectroscopy \cite{Wense4}. Such an experiment would also provide the first direct laser excitation of a nuclear transition and could be used for the development of a CEMS-based solid-state nuclear clock \cite{Wense1}.\\[0.2cm]
Experiments that are aiming for a measurement of the hyperfine shift of $^{229\text{m}}$Th based on the different spins of the nuclear ground and excited states are currently conducted at the University of Jyväskylä, Finland \cite{Sonnenschein}, by the PTB in Germany in collaboration with our group and at the KU Leuven, Belgium. A successful experiment would allow to employ the double-resonance method proposed in Ref.~\cite{Peik} for a secure identification for the isomer's population in the nuclear clock concept\footnote{Addendum: A successful observation of the $^{229\text{m}}$Th hyperfine structure was reported in 2018 \cite{Thielking}.}. The recently observed unexpected short isomeric lifetime in $^{229}$Th$^{1+}$ of below 10~ms is an important information for such experiments, as it will most likely impose the requirement to consider thorium ions of higher charge state \cite{Seiferle1}.\\[0.2cm]
When combined, the results of these experiments will pave the way for the direct laser excitation of $^{229\text{m}}$Th ions in a Paul trap and in this way for the development of a single-ion nuclear clock. Preparations are currently ongoing in the US \cite{Campbell3} in Russia \cite{Borisyuk} and in Germany \cite{Zimmermann}.\\[0.2cm]
For the development of a solid-state nuclear clock based on Mössbauer spectroscopy of $^{229}$Th as proposed in Refs.~\cite{Peik,Rellergert}, further challenges have to be solved. The central problem here is that chemical conditions have to be found that allow for a significant suppression of the IC decay channel. While this should theoretically be possible by using a material providing a band gap that is larger than the isomeric energy as a host material, experiments have so far failed to generate a secure suppression of the IC decay channel and the isomeric state was found to be extraordinary sensitive to any kind or surface coupling \cite{Wense1}. For this reason it is currently not clear if chemical conditions can be generated that will allow for the development of a solid-state nuclear clock based on the $^{229}$Th radiative decay. The development of a solid-state nuclear clock which uses the IC decay to probe the nuclear excitation (CEMS-based nuclear clock) could provide an alternative.\\[0.2cm]
A potentially promising way of future research could also be to probe solid noble gases like frozen argon as a type of host material. It is possible to probe the isomeric IC lifetime on different substrates to test if a significant lifetime prolongation can be achieved. This path of research might be favorable compared to experiments which rely on the observation of a potential radiative decay channel. In order to give any result, the latter experiments would require a suppression of the IC decay channel by up to 9 orders of magnitude, which can be considered as extremely challenging.

\subsection{Prospects for a $^{229}$Th frequency standard}
A $^{229}$Th nuclear clock is expected to outperform even the best atomic-shell based clocks due to conceptual advantages \cite{Peik,Campbell}. An estimate of the total uncertainty budget results in an expected value approaching $10^{-19}$, which is by about an order of magnitude better than the best atomic clocks currently in operation.
At the same time, a solid-state nuclear clock, although being less accurate, could provide a very compact and robust device for time measurement and thus being of significant practical use.\\[0.2cm]
However, one should expect that also the atomic clock technology will continue to improve and both technologies may evolve towards some ultimate accuracy limit. To give an example: Today's most accurate clocks already allow to measure gravitational shifts that correspond to a few centimeters difference in altitude due to general relativistic effects. When pushing the limits by further two orders of magnitude, gravitational shifts in the sub-millimeter region would become measurable. This distance is, however, shorter than the dimensions of the atoms in an optical lattice clock, so that the technology is approaching a limit. Further, any useful comparison of two clocks would require the definition of equal-time hypersurfaces around the globe. For these reasons, a future generation of ultra-precise clocks could become sensitive devices for gravitational shift measurements and geodesy instead of instruments of pure time measurement \cite{Safronova}. Also more precise tests of general relativity could be performed. The comparison of a nuclear clock with an atomic clock could provide significantly more stringent tests for potential time variations of fundamental constants \cite{Flambaum}. Of course there is the hope that $^{229}$Th will just be a first candidate for an ultra-precise nuclear clock and that, with ever evolving frequency-comb technology, a whole new generation of clocks based on nuclear transitions could be developed.

\subsection{Beyond the present limits}
The limiting factor for the development of a broad variety of nuclear clocks is the availability of frequency combs in the x-ray region with energies of a few keV and beyond. As the frequency-comb technology is a young and developing field of research, one might expect that at some point in the future the technology will be available to use nuclear excited states other than $^{229\text{m}}$Th for time measurement. Current directions of research are, e.g., high-harmonic generation \cite{Cingoez} and x-ray pulse shaping \cite{Cavaletto}.\\[0.2cm]
It is worth asking which nuclear transitions would be good clock candidates, assuming that a future frequency-comb providing up to 20~keV energy with reasonable intensity would be available. For the development of a single-ion nuclear clock, it would be possible to suppress any IC decay branch by choosing a sufficiently high charge state of the ion. This leads to a prolongation of the isomeric lifetime and correspondingly to a linewidth reduction. The expected stabilities of the single-ion nuclear clocks can be compared with the help of the Allan deviation, which is given for the case of a quantum-projection-noise (QPN) limited clock by Eq.~(\ref{Allan1}). In reality, further limitations due to different forms of noise will occur, which will not be discussed here. We use $T=1/(2\pi\Delta\nu)$ as an estimate for the interrogation cycle. In case that the nuclear transition has a lifetime significantly longer than 1~s, the stability will not be limited by the natural linewidth, but instead by the laser used for interrogation. We take this into account by assuming a natural linewidth of 1~Hz in these cases. A list of potential candidates for a single-ion nuclear clock is given in table \ref{tab:1}. All nuclear isomeric states below 20~keV excitation energy (for which sufficient data is provided in the data base) are listed, in case that a suppression of the IC decay channel would potentially lead to a half-life prolongation into the ms region or above. Only transitions providing a multipolarity of $L=2$ or smaller are listed, as otherwise no significant laser-nucleus interaction can be expected in the given energy range. The table is based on Ref.~\cite{NNDC} and further nuclear clock candidates may potentially exist, for which not sufficient data is listed in the data base.\\[0.2cm]
\begin{table*}
\caption{List of candidates for a single-ion nuclear clock with established parameters (based on Ref.~\cite{NNDC}). From left to right it is listed: (1) The energy of the excited state, (2) the nucleus together with the charge state required to suppress IC, (3) the half-life of the nucleus in its ground state, (4) the multipolarity of the transition, (5) the excited state's half-life under IC, (6) the IC coefficient, (7) the expected half-life if IC is suppressed, (8) the quality factor of the corresponding resonator, (9) the minimum laser intensity required to drive Rabi oscillations of the transition, (10) the Allan deviation (assuming for simplicity limitation due to quantum projection noise and a minimum linewidth of 1~Hz that can effectively be used).}
\label{tab:1}       
\begin{tabular}{llllllllll}
\hline\noalign{\smallskip}
E [keV]& Ion & $t_{1/2}$ gnd. & Mult. & $t_{1/2}$(ic)  & $\alpha_\text{ic}$ & $t_{1/2}$($\gamma$) & $Q(\gamma)$ & $I$ [W/cm$^2$] & $\sigma_\text{y}\cdot\sqrt{\tau}$\\
\noalign{\smallskip}\hline\noalign{\smallskip}
0.0078 & $^{229}$Th$^{2+}$ & 7932 y & M1 & 7 $\mu$s & $1\cdot10^9$ & 1.9 h & $1.2\cdot10^{20}$ & $1.2\cdot10^{-2}$ & $2.1\cdot10^{-16}$ \\
2.3 & $^{205}$Pb$^{46+}$ & $1.7\cdot10^{7}$ y & E2 & 24.2 $\mu$s & $4.4\cdot10^8$ & 3.0 h & $5.4\cdot10^{22}$ & $4.8\cdot10^5$ & $7.2\cdot10^{-19}$ \\
4.5 & $^{219}$Rn$^{58+}$ & 3.96 s & E2 & 15.4 ns & $4.3\cdot10^6$ & 67 ms & $6.5\cdot10^{17}$& $6.2\cdot10^1$ & $4.7\cdot10^{-19}$\\
5.8 & $^{185}$Ir$^{58+}$ & 14.4 h & E2 & 5 ns & $1.2\cdot10^6$ & 6.1 ms & $7.6\cdot10^{16}$ & $1.5\cdot10^3$ & $1.2\cdot10^{-18}$\\
6.5 & $^{129}$Cs$^{45+}$ & 32 h & E2 & 72 ns & $4.3\cdot10^5$ & 31 ms & $4.4\cdot10^{17}$ & $4.0\cdot10^2$ & $4.8\cdot10^{-19}$\\
8.8 & $^{249}$Bk$^{71+}$ & 330 d & M2 & 0.3 ms & $3\cdot10^6$ & 15 min & $1.7\cdot10^{22}$ &  $2.2\cdot10^6$  & $1.9\cdot10^{-19}$\\ 
10.8 & $^{124}$Sb$^{49+}$ & 60.2 d & M2 & 93 s & $2.2\cdot10^4$ & 23.1 d & $4.8\cdot10^{25}$ & $9.6\cdot10^9$ & $1.4\cdot10^{-19}$\\
12.4 & $^{45}$Sc$^{21+}$ & stable & M2 & 318 ms & 632 & 201 s & $5.5\cdot10^{21}$ & $1.4\cdot10^6$ & $1.3\cdot10^{-19}$\\
13.3 & $^{73}$Ge$^{30+}$ & stable & E2 & 2.9 $\mu$s & 1120 & 3.3 ms & $9.5\cdot10^{16}$ & $3.2\cdot10^4$ & $7.2\cdot10^{-19}$ \\
16.3 & $^{182}$Ta$^{67+}$ & 114.7 d & M2 & 283 ms & $4.3\cdot10^4$ & 3.4 h & $4.3\cdot10^{23}$ & $1.9\cdot10^8$ & $1.0\cdot10^{-19}$\\
18.2 & $^{67}$Ge$^{32+}$ & 18.9 min & E2 & 13.7 $\mu$s & 364 & 5 ms & $2.0\cdot10^{17}$ & $5.4\cdot10^{4}$ & $4.3\cdot10^{-19}$\\

\noalign{\smallskip}\hline
\end{tabular}
\end{table*}
The limiting condition for driving Rabi oscillations can be approximated as\footnote{Addendum: A factor four has been introduced as a correction compared to the previous version.} $\Omega\gtrapprox \text{max}(\Gamma_\text{tot},\Gamma_\text{L})$ with $\Omega$ the angular Rabi frequency, $\Gamma_\text{tot}=(1+\alpha_\text{ic})\Gamma_\gamma$ the total decay rate of the transition, $\Gamma_\gamma=\text{ln}(2)/t_{1/2}(\gamma)$ the radiative decay rate of the transition and $\Gamma_\text{L}$ the bandwidth of the laser light used for nuclear excitation. Using the definition of the Rabi frequency as $\Omega=\sqrt{2\pi c^2I\Gamma_\gamma/(\hbar\omega^3)}$ one obtains the condition
\begin{equation}
\sqrt{\frac{2\pi c^2 I\Gamma_\gamma}{\hbar\omega^3}}\gtrapprox \text{max}(\Gamma_\text{tot},\Gamma_\text{L}),
\end{equation}
with $I$ the laser intensity,  For $\Gamma_\text{L}>\Gamma_\text{tot}$ this leads to 
\begin{equation}
I\gtrapprox\frac{\hbar\omega^3\Gamma_\text{L}^2}{2\pi c^2\Gamma_\gamma}.
\end{equation}
In case of $\Gamma_\text{tot}>\Gamma_\text{L}$ one obtains
\begin{equation}
I\gtrapprox\frac{\hbar\omega^3(1+\alpha_\text{ic})^2\Gamma_\gamma}{2\pi c^2}.
\end{equation}
These equations were used to calculate the laser intensities required to drive nuclear Rabi oscillations for the case that a single mode of a frequency comb with $\Gamma_L=2\pi$ bandwidth was used for driving the transition. Note, that for all single-ion nuclear clocks it was assumed that the IC decay branch is completely suppressed ($\alpha_\text{ic}=0)$, which can be considered as an idealistic scenario, as even in case of successful suppression of IC, bound internal conversion processes can be assumed to be present.\\[0.2cm] 
For single-ion nuclear clocks, also nuclei of relatively short half-lives can be taken into consideration, as the number of required ions could be kept small and a Paul trap could be frequently loaded. However, operating a nuclear clock based on short-lived nuclei might require significant effort. For this reason the only candidates, which might be practical for operation, would be (except $^{229}$Th): $^{205}$Pb$^{46+}$ (2.3~keV, 3~h, E2), $^{45}$Sc$^{21+}$ (12.4~keV, 201~s, M2) and $^{73}$Ge$^{30+}$ (13.3~keV, 3.3~ms, E2).\\[0.2cm]
\begin{table*}
\caption{List of candidates for a Mössbauer nuclear clock (based on Ref.~\cite{NNDC}). From left to right it is listed: (1) The energy of the excited state, (2) the nucleus, (3) the half-life of the nucleus in its ground state, (4) the multipolarity of the transition, (5) the isomer's half-life under IC, (6) the IC coefficient, (7) the quality factor of the corresponding resonator if IC is not suppressed, (8) the number of excited nuclei per second for the considered irradiation scheme, (9) the expected Allan deviation calculated based on Eq.~(\ref{Allan2}).}
\label{tab:0}       
\begin{tabular}{lllllllll}
\hline\noalign{\smallskip}
E [keV]& Nucleus & $t_{1/2}$ gnd. & Mult. & $t_{1/2}$(ic)  & $\alpha_{\text{ic}}$   & $Q(\text{ic})$ & $N_\text{exc}/s$  &$\sigma_\text{y}\cdot\sqrt{\tau}$\\
\noalign{\smallskip}\hline\noalign{\smallskip}
0.0078 & $^{229}$Th & 7932 y & M1 & 7 $\mu$s  & $1\cdot10^9$  & $1.2\cdot10^{11}$ & $2.5\cdot10^7$ & $3.7\cdot10^{-15}$ \\
1.6 & $^{201}$Hg & stable & M1 & 81 ns & $4.7\cdot10^4$ & $2.8\cdot10^{11}$ & $6.2\cdot10^4$ & $3.1\cdot10^{-14}$ \\
2.3 & $^{205}$Pb & $1.7\cdot10^7$ y & E2 & 24 $\mu$s & $4.4\cdot10^8$ & $1.2\cdot10^{14}$ & 2.2 & $1.2\cdot10^{-14}$\\
6.2 & $^{181}$Ta & stable & E1 & 6.1 $\mu$s & 70.5 & $8.3\cdot10^{13}$ & $7.1\cdot10^5$ & $3.2\cdot10^{-17}$ \\
8.4 & $^{169}$Tm & stable & M1 & 4.1 ns & 263 & $7.5\cdot10^{10}$ & $7.7\cdot10^4$ & $1.1\cdot10^{-13}$ \\
9.4 & $^{83}$Kr & stable & M1  & 157 ns & 17.1 & $3.2\cdot10^{12}$ & $8.0\cdot10^5$ & $7.7\cdot10^{-16}$ \\
9.8 & $^{187}$Os & stable & M1 & 2.4 ns & 280 & $5.2\cdot10^{10}$ & $ 4.5\cdot10^4$ & $ 2.0\cdot10^{-13}$ \\
10.6 & $^{137}$La & $6\cdot10^4$ y & M1 & 89 ns & 117.6 & $2.1\cdot10^{12}$ & $8.5\cdot10^4$ & $3.7\cdot10^{-15}$ \\
12.4 & $^{45}$Sc & stable & M2 & 318 ms & 632  & $8.6\cdot10^{18}$   & $9.9\cdot10^3$  & $2.6\cdot10^{-21}$\\
13.3 & $^{73}$Ge & stable & E2 & 2.9 $\mu$s & 1120  &  $8.4\cdot10^{13}$  & $4.6\cdot10^3$ & $3.9\cdot10^{-16}$\\
14.4 & $^{57}$Fe & stable & M1 & 98 ns & 8.6  &  $3.1\cdot10^{12}$  & $4.2\cdot10^5$ & $1.1\cdot10^{-15}$\\
\noalign{\smallskip}\hline
\end{tabular}
\end{table*}
For nuclear clocks based on Mössbauer spectroscopy the discussion is slightly complicated by the fact that the Ramsey interrogation scheme, which is the basis for the derivation of the QPN limited Allan deviation in Eq.~(\ref{Allan1}), cannot be applied \cite{Kazakov1}. Instead, a fluorescence spectroscopy interrogation scheme has been proposed, which, in the realistic limit of a short interrogation time and for a weak laser field, results in a shot-noise limited Allan deviation given as \cite{Kazakov1}
\begin{equation}
\sigma_y(\tau)=\frac{9}{2}\frac{1}{\omega T}\sqrt{\frac{1}{RN\tau}}.
\end{equation}
Here $R$ denotes the nuclear excitation rate and all other variables are defined as previously. We have assumed that the decay rate of the coherences (see Ref.~\cite{Kazakov1}) can be approximated as $\Gamma_\text{tot}/2$. Further, the number of effectively detected decay events was set to be equal to the number of excited nuclei. This corresponds to the assumption of 100\% detection efficiency and that the detection is performed in the IC decay channel (being the dominant decay channel for all transitions), thus considering only nuclear clocks based on laser conversion-electron Mössbauer spectroscopy (CEMS) \cite{Wense1}. In a linear approximation the excitation rate $R$ is given as\footnote{Addendum: A factor of $2/\pi$ has been introduced as a correction compared to the previous version.}
\begin{equation}
R=\frac{2\pi c^2 I\Gamma_\gamma}{\hbar \omega^3 \Gamma_\text{L}},
\end{equation}
Inserting this into the expression for the Allan deviation leads to
\begin{equation}
\label{Allan2}
\sigma_y(\tau)\approx \sqrt{\frac{\hbar\omega \Gamma_\text{L}}{I c^2 \Gamma_\gamma N T^2 \tau}}.
\end{equation}
For the following comparison it is assumed that a number of $N=10^{14}$ nuclei located on an area of 1~mm$^2$ is irradiated by a single mode of a frequency comb of a bandwidth smaller or equal to the nuclear transition linewidth ($\Gamma_\text{L}=\Gamma_\text{tot}$) and an intensity of $I=10^{-5}$~W/cm$^2$. In this case the total number of excited nuclei per second is estimated as $N_\text{exc}\approx RN/2$. Only transitions with half-lives of longer than 1~ns are considered in an energy range of up to 20~keV, for which the ground state is either stable or sufficiently long-lived in order to handle the required quantity of material. Further, only transitions with a multipolarity of $L=2$ or smaller are considered, as higher multipolarities do not allow for an efficient laser-nucleus coupling. The resulting transitions are listed in Tab.~\ref{tab:0}, the 12.4~keV transition of $^{45}$Sc provides the best estimated stability.\\[0.2cm]
Rather surprisingly, even a CEMS nuclear clock based on $^{229\text{m}}$Th could have the potential to provide very high stability of $\sigma_\text{y}\sqrt{\tau}=3.7\cdot10^{-15}$ despite the IC-induced line-broadening to 15.9~kHz. The CEMS-based nuclear clock should not be confused with the crystal-lattice nuclear clock based on $^{229}$Th as proposed in Refs.~\cite{Peik,Rellergert}, for which it is the concept to suppress the IC decay channel.\\[0.2cm]
It can be inferred, that an exciting new area of research could open up in case that x-ray frequency combs will become available in the future, potentially leading to the development of a variety of different nuclear clocks. Due to its extraordinary low excitation energy, $^{229\text{m}}$Th will be the first candidate for the development of a nuclear clock for which all required technology is already available by today.

\begin{acknowledgements}
This work was supported by the European Union's Horizon 2020 research and innovation programme under grant agreement No.~664732 "nuClock", by the DFG grant Th956/3-1 and by the LMU department of Medical Physics via the Maier-Leibnitz Laboratory. L.v.d.Wense would like to thank the organizers of the conference within the frame of the jubelee celebrating "175 years of the D.I. Mendeleyev Institute for Metrology (VNIIM) and National Measurement System" in St. Petersburg for the invitation.
\end{acknowledgements}


\begin{thebibliography}{spmpsci}
%
%

\bibitem{Higgins} K. Higgins, D. Miner, C.N. Smith \& D.B. Sullivan, {\it A Walk Through Time} (version 1.2.1) (2004). Online available at: {\it http://physics.nist.gov/time} [2010, July 12] National Institute of Standards and Technology, Gaithersburg, MD.
\bibitem{Bennet} M. Bennet et al., {\it Huygens'clocks}, Proceedings of the Royal Society of London A \textbf{458}, 563-579 (2002).
\bibitem{Sorge} F. Sorge, M. Cammalleri \& G. Genchi, {\it On the birth and growth of pendulum clocks in the early modern era}, in: Essays on the history of mechanical engineering, 273-290 Springer (2016).
\bibitem{Gould} R.T. Gould, {\it The marine chronometer: its history and development}, J.D. Potter (1923).
\bibitem{Bosschieter} J.E. Bosschieter, {\it Shortt's free pendulum}, A History of the Evolution of Electric Clocks. Online available at: {\it http://www.electric-clocks.eu/clocks/en/page10.htm} [2017, September 3rd]
\bibitem{Marrison} W.A. Marrison, {\it The evolution of the quartz crystal clock}, The Bell System Technical Journal \textbf{27}, 510-588 (1948).
\bibitem{Lyons} H. Lyons, {\it The atomic clock}, Instruments Vol. \textbf{22}, 133-135 (1949).
\bibitem{Forman} P. Forman, {\it Atomichron: The atomic clock from concept to commercial product}, IEEE Ultrasonics, Ferroelectrics and Frequency Control Society (1998).
\bibitem{Essen} L. Essen \& J.V.L. Parry, {\it An atomic standard of frequency and time interval: A caesium resonator}, Nature \textbf{176}, 280-282 (1955).
\bibitem{Ramsey} N.F. Ramsey, {\it History of Atomic clocks}, Journal of Research of the National Bureau of Standards \textbf{88}, 301-318 (1983).
\bibitem{Wynands} R. Wynands \& S. Weyers, {\it Atomic fountain clocks}, Metrologica \textbf{42}, 64-79 (2005).
\bibitem{Udem} Th. Udem, R. Holzwarth \& T.W. Hänsch, {\it Optical frequency metrology}, Nature \textbf{416}, 233-237 (2002).
\bibitem{Diddams} S.A. Diddams et al., {\it An optical clock based on a single trapped $^{199}$Hg$^+$ ion}, Science \textbf{293}, 825-828 (2001).
\bibitem{Ludlow} A.D. Ludlow et al., {\it Optical atomic clocks}, Rev. Mod. Phys. \textbf{87}, 637-699 (2015).
\bibitem{Rosenband} T. Rosenband et al., {\it Frequency Ratio of Al$^+$ and Hg$^+$ Single-ion optical clocks; Metrology at the 17th decimal place}, Science \textbf{319}, 1808-1811 (2008).
\bibitem{Huntemann} N. Huntemann et al., {\it Single-ion atomic clock with $3\cdot10^{-18}$ systematic uncertainty}, Phys. Rev. Lett. \textbf{116}, 063001 (2016).
\bibitem{Bloom} B.J. Bloom et al., {\it An optical lattice clock with accuracy and stability at the $10^{-18}$ level}, Nature \textbf{506}, 71-75 (2014).
\bibitem{Nicholson} T.L. Nicholson et al., {\it Systematic evaluation of an atomic clock at $2\cdot10^{-18}$ total uncertainty}, Nature Communications (2015).
\bibitem{Mcgrew} W.F. McGrew et al., {\it Atomic clock performance beyond the geodetic limit}, arXiv: 1807.11282 (2018).

\bibitem{Peik2} E. Peik \& M. Okhapkin, {\it Nuclear clocks based on resonant excitation of $\gamma$-transitions}, Comptes Rendus Physique \textbf{16}, 516-523 (2015).
\bibitem{Kroger} L.A. Kroger \& C.W. Reich, {\it Features of the low energy level scheme of $^{229}$Th as observed in the $\alpha$ decay of $^{233}$U}, Nucl. Phys. A \textbf{259}, 29 (1976).
\bibitem{Reich} C.W. Reich \& R. Helmer, {\it Energy separation of the doublet of intrinsic states at the ground state of $^{229}$Th}, Phys. Rev. Lett. \textbf{64}, 271 (1990).
\bibitem{Helmer} R. Helmer \& C.W. Reich, {\it An excited state of $^{229}$Th at 3.5~eV}, Phys. Rev. C \textbf{49}, 1845 (1994).
\bibitem{Beck1} B.R. Beck et al., {\it Energy splitting of the ground-state doublet in the nucleus $^{229}$Th}, Phys. Rev. Lett. \textbf{109}, 142501 (2007).
\bibitem{Beck2} B.R. Beck et al., {\it Improved value for the energy splitting of the ground-state doublet in the nucleus $^{229\text{m}}$Th}, LLNL-PROC-415170 (2009).
\bibitem{Karpeshin} F.F. Karpeshin \& M.B. Trzhaskovskaya, {\it Impact of the electron environment on the lifetime of the $^{229}$Th$^\text{m}$ low-lying isomer}, Phys. Rev. C \textbf{76}, 054313 (2007).
\bibitem{Wense1} L. von der Wense, {\it On the direct detection of $^{229\text{m}}$Th}, Ph.D. thesis, Ludwig-Maximilians-Universität München, Germany (2016). Online available at: {\it https://edoc.ub.uni-muenchen.de/20492/7/ Wense\_Lars\_von\_der.pdf}
\bibitem{Vorykhalov} O.V. Vorykhalov \& V.V. Koltsov, {\it Search for an isomeric transition of energy below 5~eV in $^{229}$Th nucleus}, Bull. Russ. Acad. Sci.: Physics \textbf{59}, 20-24 (1995).
\bibitem{Strizhov} V.F. Strizhov \& E.V. Tkalya, {\it Decay channel of low-lying isomer state of the $^{229}$Th nucleus. Possibilities of experimental investigation}, Sov. Phys. JETP \textbf{72}, 387 (1991).
\bibitem{Tkalya1} E.V. Tkalya, V.O. Varlamov, V.V. Lomonosov, S.A. Nikulin, {\it Processes of the nuclear isomer $^{229\text{m}}$Th(3/2$^+$, 3.5$\pm$1.0~eV) resonant excitation by optical photons}, Phys. Scripta \textbf{53}, 296-299 (1996).
\bibitem{Tkalya2} E.V. Tkalya, A.N. Zherikin \& V.I. Zhudov, {\it Decay of the low-energy nuclear isomer $^{229}$Th$^\text{m}$(3/2$^+$,3.5$\pm$1.0~eV) in solids (dielectrics and metals): a new scheme of experimental research}, Phys. Rev. C \textbf{61}, 064308 (2000).
\bibitem{Peik} E. Peik \& C. Tamm, {\it Nuclear laser spectroscopy of the 3.5~eV transition in $^{229}$Th}, Europhys. Lett. \textbf{61}, 181-186 (2003).
\bibitem{Minkov} N. Minkov \& A. Pálffy, {\it Reduced transition probabilities for the gamma decay of the $7.8$~eV isomer in $^{229}$Th}, Phys. Rev. Lett. \textbf{118}, 212501 (2017).
\bibitem{Tkalya3} E.V. Tkalya, C. Schneider, J. Jeet \& E.R. Hudson, {\it Radiative lifetime and energy of the low-energy isomeric level in $^{229}$Th}, Phys. Rev. C \textbf{92}, 054324 (2015).
\bibitem{Campbell} C.J. Campbell, A.G. Radnaev, A. Kuzmich, V.A. Dzuba, V.V. Flambaum \& A. Derevianko, {\it Single-ion nuclear clock for metrology at the 19th decimal place}, Phys. Rev. Lett. \textbf{108}, 120802 (2012).
\bibitem{Campbell3} C.J. Campbell, A.G. Radnaev \& A. Kuzmich, {Wigner crystals of $^{229}$Th for optical excitation of the nuclear isomer}, Phys. Rev. Lett. \textbf{106}, 223001 (2011).
\bibitem{Zimmermann} K. Zimmermann, {\it Experiments towards optical nuclear spectroscopy with thorium-229}, Ph.D. thesis, University of Hannover, Germany (2010).
\bibitem{Borisyuk} P.V. Borisyuk et al., {\it Trapping, retention and laser cooling of Th$^{3+}$ ions in a multisection linear quadrupole trap}, Quantum Electronics \textbf{47}, 406-411 (2017).
\bibitem{Rellergert} W.G. Rellergert et al., {\it Constraining the evolution of the fundamental constants with a solid-state optical frequency reference based on the $^{229}$Th nucleus}, Phys. Rev. Lett. \textbf{104}, 200802 (2010).
\bibitem{Kazakov1} G.A. Kazakov et al., {\it Performance of a $^{229}$Thorium solid-state nuclear clock}, New Jour. Phys. \textbf{14}, 083019 (2012).

\bibitem{Swanberg} E. Swanberg, {\it Searching for the decay of $^{229\text{m}}$Th}, Ph.D. thesis, University of California, Berkeley (2012).
\bibitem{Zhao} X. Zhao et al., {\it Observation of the deexcitation of the $^{229\text{m}}$Th nuclear isomer}, Phys. Rev. Lett. \textbf{109}, 160801 (2012).
\bibitem{Peik3} E. Peik \& K. Zimmermann, {\it Comment on "Observation of the deexcitation of the $^{229\text{m}}$Th nuclear isomer"}, Phys. Rev. Lett. \textbf{111}, 018901 (2013).
\bibitem{Wense2} L. von der Wense et al., {\it Towards a direct transition energy measurement of the lowest nuclear excitation in $^{229}$Th}, JINST \textbf{8}, P03005 (2013).
\bibitem{Hehlen} M.P. Hehlen et al., {\it Optical spectroscopy of an atomic nucleus: Progress toward direct observation of the $^{229}$Th isomer transition}, J. Lumin. \textbf{133}, 91-95 (2013).
\bibitem{Stellmer3} S. Stellmer et al., {\it Feasibility study of measuring the $^{229}$Th nuclear isomer transition with $^{233}$U-doped crystals}, Phys. Rev. C \textbf{94}, 014302 (2016).
\bibitem{Duppen} P. van Duppen et al., {\it Characterization of the low-energy $^{229\text{m}}$Th isomer}, Letter of intent to the ISOLDE and neutron time-of-flight committee (2017). Online available at: {\it cds.cern.ch/record/2266840}

\bibitem{Porsev} S.G. Porsev, V.V. Flambaum, E. Peik \& Chr. Tamm, {\it Excitation of the isomeric $^{229\text{m}}$Th nuclear state via an electronic bridge process in $^{229}$Th$^{1+}$}, Phys. Rev. Lett. \textbf{105}, 182501 (2010). 
\bibitem{Herrera} O.A. Herrera-Sancho, {\it Laser excitation of 8-eV electronic states in Th$^+$: A first pillar of the electronic bridge toward excitation of the Th-229 nucleus}, Ph.D. thesis, Univ. Hannover, Germany (2012).
\bibitem{Campbell2} C.J. Campbell et al., {Multiply charged thorium crystals for nuclear laser spectroscopy}, Phys. Rev. Lett. \textbf{102}, 233004 (2009).

\bibitem{Jeet} J. Jeet et al., {\it Results of a direct search using synchrotron radiation for the low-energy $^{229}$Th nuclear isomeric transition}, Phys. Rev. Lett. \textbf{114}, 253001 (2015).
\bibitem{Stellmer} S. Stellmer, M. Schreitl \& T. Schumm, {\it Radioluminescence and photoluminescence of Th:CaF$_2$ crystals}, Scientific Reports \textbf{5}, 15580 (2015).
\bibitem{Stellmer4} S. Stellmer et al., {\it Attempt to optically excite the nuclear isomer in $^{229}$Th}, Phys. Rev. A \textbf{97}, 062506 (2018).
\bibitem{Yamaguchi} A. Yamaguchi et al., {\it Experimental search for the low-energy nuclear transition in $^{229}$Th with undulator radiation}, New J. Phys. \textbf{17} 053053 (2015).
\bibitem{Yoshimi} A. Yoshimi et al., {\it Nuclear resonant scattering experiment with fast time response: new scheme for observation of $^{229\text{m}}$Th radiative decay}, Phys. Rev. C \textbf{97}, 024607 (2018).
\bibitem{Gangrsky} Yu.P. Gangrsky et al., {\it Search for light radiation in decay of $^{229}$Th isomer with anomalously low excitation energy}, Bull. Rus. Acad. Sci. Phys. \textbf{69} 1857 (2005).
\bibitem{Stellmer5} S. Stellmer et al., {\it Toward an energy measurement of the internal conversion electron in the deexcitation of the $^{229}$Th isomer}, Phys. Rev. C \textbf{98}, 014317 (2018).
\bibitem{Wense3} L. von der Wense et al., {\it Direct detection of the $^{229}$Th nuclear clock transition}, Nature \textbf{533}, 47-51 (2016).
\bibitem{Seiferle1} B. Seiferle, L. von der Wense \& P.G. Thirolf, {\it Lifetime measurement of the $^{229}$Th nuclear isomer}, Phys. Rev. Lett. \textbf{118}, 042501 (2017).
\bibitem{Seiferle2} B. Seiferle, L. von der Wense \& P.G. Thirolf, {\it Feasibility study of internal conversion electron spectroscopy of $^{229\text{m}}$Th}, Eur. Phys. J. A \textbf{53}, 108 (2017).
\bibitem{Ponce} F. Ponce, {\it High accuracy measurement of the nuclear decay of U-235m and search for the nuclear decay of Th-229m}, Ph.D. thesis, University of California, USA (2017).
\bibitem{Varlamov} V.O. Varlamov et al., {\it Excitation of a $^{229\text{m}}$Th ($(3/2)^+$, 3.5~eV) isomer by surface plasmons}, Phys. Doklady \textbf{41}, 47 (1996).
\bibitem{Wense4} L. von der Wense et al., {\it Laser excitation scheme for $^{229\text{m}}$Th}, Phys. Rev. Lett. \textbf{119}, 132503 (2017).
\bibitem{Kazakov} G.A. Kazakov et al., {\it Prospects for measuring the $^{229}$Th isomer energy using a metallic magnetic microcalorimeter}, Nucl. Instrum. Methods A \textbf{735}, 229-239 (2014).
\bibitem{Schneider} P. Schneider, {\it Spektroskopische Messungen an Thorium-229 mit einem Detektor-Array aus metallischen magnetischen Kalorimetern}, Master thesis, Ruprecht-Karls-Universität Heidelberg, Germany (2016).
\bibitem{Palffy} A. Pálffy et al., {\it Isomer triggering via nuclear excitation by electron capture}, Phys. Rev. Lett. \textbf{99}, 172502 (2007).
\bibitem{Brandau} C. Brandau et al., {\it Probing nuclear properties by resonant atomic collisions between electrons and ions}, Phys. Scr. \textbf{T156}, 014050 (2013).
\bibitem{Ma} X. Ma et al., {\it Proposal for precision determination of 7.8~eV isomeric state in $^{229}$Th at heavy ion storage ring}, Phys. Scr. \textbf{T166}, 014012 (2015).
\bibitem{Liao} W.T. Liao \& A. Pálffy, {\it Optomechanically induced transparency of x-rays via optical control}, Scientific Reports \textbf{7}, 321 (2017).
\bibitem{Beloy} K. Beloy, {\it Hyperfine structure in $^{229\text{g}}$Th$^{3+}$ as a probe of the $^{229\text{g}}$Th $\rightarrow$ $^{229\text{m}}$Th nuclear excitation energy}, Phys. Rev. Lett. \textbf{112}, 062503 (2014).
\bibitem{Thielking} J. Thielking et al., {\it Laser spectroscopic characterization of the nuclear-clock isomer $^{229\text{m}}$Th}, Nature \textbf{556}, 321 (2018).
\bibitem{Sonnenschein} V. Sonnenschein et al., {\it The search for the existence of $^{229\text{m}}$Th at IGISOL}, Eur. Phys. J. A \textbf{48}, 52 (2012).
\bibitem{Safronova} M. Safronova, {\it Elusive transition spotted in thorium}, Nature \textbf{533}, 44-45 (2016).
\bibitem{Flambaum} V.V. Flambaum, {\it Enhanced effect of temporal variation of the fine structure constant and the strong interaction in $^{229}$Th}, Phys. Rev. Lett. \textbf{97}, 092502 (2006).
\bibitem{Cingoez} A. Cingöz et al., {\it Direct frequency comb spectroscopy in the extreme ultraviolet}, Nature \textbf{482}, 68-71 (2012).
\bibitem{Cavaletto} S.M. Cavaletto et al., {\it Broadband high-resolution x-ray frequency combs}, Nature Photonics \textbf{8}, 520-523 (2014).

\bibitem{NNDC} NNDC interactive chart of nuclides, Online available at: {\it https://www.nndc.bnl.gov/chart} [2017, September 3rd], Brookhaven National Laboratory, Brookhaven.

\end{thebibliography}


\end{document}